\documentclass[twocolumn]{article}
\usepackage[T1]{fontenc}
\usepackage[utf8]{inputenc}
\usepackage{mathptmx} 
\usepackage[margin=0.75in]{geometry}
\usepackage[affil-it]{authblk}
\usepackage{amsfonts, amsmath, amssymb}
\usepackage{hyperref}
\usepackage{chemformula}
\usepackage{booktabs}
\usepackage{graphicx}
\usepackage[style=nature]{biblatex} 
\usepackage[raggedright]{titlesec}
\usepackage[labelfont=bf,textfont=md]{caption}
\usepackage{algorithm, algpseudocode}
\usepackage[final]{changes}

\title{Score-based denoising for atomic structure identification}

\author[1]{Tim Hsu\thanks{hsu16@llnl.gov}}
\author[1]{Babak Sadigh}
\author[1]{Nicolas Bertin}
\author[1]{Cheol Woo Park}
\author[2]{James Chapman}
\author[1]{Vasily Bulatov}
\author[1]{Fei Zhou\thanks{zhou6@llnl.gov}}

\affil[1]{Lawrence Livermore National Laboratory, Livermore, CA, United States}
\affil[2]{Boston University, Boston, MA, United States}

\date{}

\begin{document}
\maketitle
\begin{abstract}

We propose an \replaced{effective}{accurate} method for removing thermal vibrations that complicate the task of analyzing complex dynamics in atomistic simulation of condensed matter. Our method iteratively subtracts thermal noises or perturbations in atomic positions using a denoising score function trained on synthetically noised but otherwise perfect crystal lattices. The resulting denoised structures clearly reveal underlying crystal order while retaining disorder associated with crystal defects. Purely geometric, agnostic to interatomic potentials, and trained without inputs from explicit simulations, our denoiser can be applied to simulation data generated from vastly different interatomic interactions. \replaced{The denoiser is shown to improve existing classification methods such as common neighbor analysis and polyhedral template matching, reaching perfect classification accuracy on a recent benchmark dataset of thermally perturbed structures up to the melting point.}{Followed by a simple phase classification tool such as the Common Neighbor Analysis, the denoiser outperforms other existing methods and reaches perfect classification accuracy on a recently proposed benchmark dataset consisting of perturbed crystal structures (DC3).} Demonstrated here in a wide variety of atomistic simulation contexts, the denoiser is general, robust, and readily extendable to delineate order from disorder in structurally and chemically complex materials.  

\end{abstract}

\section{Introduction}
In molecular dynamics (MD) of condensed matter, characterization methods for the simulated atomic configurations aim to unravel meaningful structural features such as crystalline phases and defects. As the simulations are typically carried out at finite temperatures, accurate characterization of structures and defects is complicated by perturbations in atomic positions induced by thermal vibrations. To this end, increasingly sophisticated methods have been proposed over the years for identification of local atomic motifs in simulated configurations \cite{Steinhardt1983PRB, Honeycutt1987JPC-CNA, Stukowski2012MSMSE-review, Tanaka2019NRP, larsen2016robust, larsen2020revisiting, ackland2006applications, lazar2015topological, nguyen2015identification}.

Existing characterization methods usually focus on either ordered crystalline phases or crystal defects. For example, the common neighbor analysis (CNA) algorithm \cite{Honeycutt1987JPC-CNA} identifies simple crystal structures such as the body-centered cubic (BCC), face-centered cubic (FCC), and hexagonal closed-packed (HCP). Other commonly used methods for structure identification include bond order analysis \cite{Steinhardt1983PRB, Lechner2008JCP}, centrosymmetry analysis \cite{Kelchner1998PRB-CSA}, adaptive template analysis \cite{Sapozhnikov2008-ATA}\added{, and polyhedral template matching (PTM) \cite{larsen2016robust}}. On the other hand, the dislocation extraction algorithm (DXA) \cite{Stukowski2010MSMSE-DXA, Stukowski2012MSMSE-DXA} identifies dislocation defects within an \textit{a priori} known ordered crystalline environment. All of the mentioned methods rely heavily on domain knowledge, physical intuition, and heuristics (for review, see for example\ \cite{Stukowski2012MSMSE-review}). As such, they are often application- and/or structure-specific, and are not always easy to generalize beyond their original scope of applicability. More recently, data-driven machine-learning (ML) approaches are being developed for performing ordered phase classification and sometimes defect detection \cite{Kim2020PCCP, Swanson2020SM, Doi2020JCP, Becker2021PRE, Leitherer2021NC, Chung2022PRM, Hernandes2022JPCM, Chapman2022nCM}, often employing existing tools such as Steinhardt order parameters \cite{Steinhardt1983PRB} for featurization. While comparatively more straightforward to develop with modern ML pipelines, these emerging methods require considerable amounts of carefully curated training data and are often informed by material-specific physics and domain knowledge which limit transferability of the trained models.

\begin{figure}
    \centering
    \includegraphics[width=0.45\textwidth]{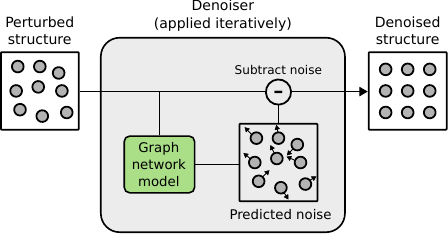}
    \caption{Our denoising graph network model predicts and subtracts thermal perturbations in atomic positions.  Denoising proceeds iteratively until the predicted noise becomes approximately zero or after a pre-defined number of iterations. In this schematic the atoms are shown in 2D Cartesian space for ease of visualization.}
    \label{fig:intro}
\end{figure}

\added{
In this work, we take a probabilistic and physics-agnostic approach by supplementing structure characterization tasks with a \textit{denoising} model. Consider a structure classifier $\mathcal{C}$ as a prototypical characterization task but other downstream tasks will also benefit. Since $\mathcal{C}(\mathbf{x})$ predicts the discrete structural type for each atom in configuration $\mathbf{x}$, there exists a finite region of perturbing displacements $\{\mathbf{\epsilon}\}$ within which $\mathcal{C}(\mathbf{x} + \mathbf{\epsilon}) \equiv \mathcal{C}(\mathbf{x})$. Classification is easy for configurations close to the center (i.e.\ ideal configurations) of their $\epsilon$-neighborhoods, while those at the boundaries pose challenges to $\mathcal{C}$. Rather than directly improving $\mathcal{C}$, we propose to first optimize any configuration $\mathbf{x}$ by bringing it towards the center of its $\epsilon$-neighborhood before classification. This amounts to an improved composite classifier $\mathcal{C} \circ \mathcal{D}$, where $\mathcal{D}$ is a denoising function to remove perturbations with respect to certain ideal reference structures. The probabilistic interpretation is that iterative applications of $\mathcal{D}$, or $\mathcal{D}^n$, gradually increase the similarity of a configuration towards the references to improve classification accuracy. In this way, the problems of classifying ordered crystalline phases and identifying disordered crystal defects can be unified, in analogy with determining order-disorder features from thermal noise. It should be noted that the denoised configurations obtained in our probabilistic approach are mathematical constructs and do not necessarily have direct physical meaning.
}

\deleted{
In this work, we consider the tasks of classifying ordered crystalline phases and identifying disordered crystal defects to be fundamentally of the same nature, akin to determining the order-disorder feature signal from thermal noise. We address this unifying problem with an iterative optimization scheme that aims to eliminate thermal noise. By minimizing thermal perturbation, our method significantly simplifies subsequent characterization of both ordered crystalline phases and crystal defects, thereby placing the two seemingly distinct tasks on an equal footing. 
}

To implement such a denoiser, we trained a graph network model, based on the equivariant NequIP architecture \cite{e3nn, Batzner2021NC-nequip}, to denoise heavily perturbed structures and reveal the underlying order-disorder (Fig.~\ref{fig:intro}). Given a distribution of pairs of noiseless ideal reference topologies (e.g., BCC, FCC, and HCP) and their noised (perturbed) counterparts, training our denoiser is equivalent to learning a score function \cite{Vincent2011scorematching}, which in this work is a gradient field in the atomic coordinate space converging towards points of maximum likelihood that correspond to the ideal reference topologies (more details in Section \ref{sec:theory}). Equipped with this theoretic knowledge, our denoiser can be considered as an iterative scheme that optimizes perturbed structures towards ideal topologies. The score function plays a central role in modern generative models such as the denoising diffusion probabilistic model (DDPM) \cite{Sohl-Dickstein2015-DPM, Ho2020-DDPM, Song2020-unified} for sampling realistic data from a high dimensional data distribution \cite{Yang2022-dpm-review}. Here it is applied for denoising rather than generative applications. \replaced{Iterative denoising with score functions allows}{Iterative applications of score functions allow} us to approach perfect identification of ordered crystalline phases in several case studies with significant improvement over exiting classifiers\replaced{}{ with a single application}. It is interesting that similar advantages of iterative use of score functions were observed in DDPM with annealed Langevin dynamics for generative purpose \cite{Sohl-Dickstein2015-DPM, Ho2020-DDPM, Song2020-unified}. Our findings support the view that iterative models break down challenging problems into smaller, manageable steps.

Compared to other existing denoising methods, e.g. energy minimization (or steepest descent mapping \cite{STILLINGER2006958}) and vibration-averaging, our method is unbiased in the sense that the ideal reference structures have the same probability, and only requires the instantaneous snapshot as the input. In contrast, energy minimization is intrinsically biased due to the use of interatomic potential that may favor certain phases over others, and vibration-averaging requires a tuned averaging window over multiple snapshots where fast processes may be overly smeared. Further, utilizing synthetically noised structures as training data, our approach does not rely on in-depth physical knowledge other than the ideal reference structures, and is a purely geometric algorithm complementary to existing physics-based techniques. As such, in contrast to data-hungry approaches, our denoiser does not need physics simulation data for training.

Prioritizing single-element systems in this work, the denoising capabilities of our model is demonstrated on several challenging applications, including identification of transient crystal phases during Cu solidification from melt, and characterization of dislocations and point defect debris in BCC Ta undergoing plastic deformation. Importantly, our denoiser does not overzealously denoise the disordered melt into ordered phases. Further, it is shown to help reveal and locate point defects, dislocations, and grain boundaries at high temperatures (approaching melting point) where again the model is observed not to denoise or rearrange crystal defects into ideal lattice motifs. Additional demonstrations on two-element \ch{SiO2} polymorphs are also provided. Besides denoising, the underyling neural network architecture of the denoiser can be extended to classify the denoised atomic environments. At this stage of development we \added{mainly} rely instead on existing methods \replaced{such as}{, mainly} CNA, \added{PTM,} and DXA to perform the final characterization. With appropriate optimization, we envision that our denoising algorithm would be a robust and highly efficient filter integrated in the workflows of massive MD simulations for the purpose of on-the-fly data compression and post-processing analyses.
\section{Results}
\added{
In our approach, denoising a thermally perturbed configuration $\mathbf{x}$ entails iteratively subtracting the noises predicted by a graph network model $\epsilon_{\theta}$ with parameters $\theta$ (Algorithm~\ref{alg:denoise}). In code implementation, $\mathbf{x}$ is expressed in terms of the atomic coordindates $\mathbf{r}$ and auxiliary information $\mathbf{z}$ regarding cell dimensions and atom types.
}
\deleted{
In our approach, denoising a thermally perturbed configuration entails iteratively subtracting the noises predicted by a machine learned graph network model (Fig.~\ref{fig:intro}):
}
\deleted{
where $\mathbf{r}$ is the atomic coordinates, $\epsilon_{\theta}$ is the graph network model with parameters $\theta$, and $\mathbf{z}$ holds auxiliary information such as the atom types and unit cell dimensions. Repeated \replaced{$\mathcal{D}^n$}{application of equation~1} amounts to finding a converged fixed point corresponding to a fully denoised structure.
}
Our denoiser is an optimization algorithm that \deleted{topologically} modifies input \replaced{noisy}{(noised)} structures towards maximal data likelihood (further explained in Section~\ref{sec:theory}, with toy visualization in Supplementary Fig.~1). By including the ideal FCC, HCP, and BCC lattices in the training data, our model attempts to evolve an input noised structure towards one of the three ideal lattices depending on which lattice type it most resembles \replaced{geometrically}{topologically}. Importantly, as demonstrated in the results that follow, our denoiser does not excessively alter the topology of disordered structures, including liquid/melt phase, point defects, dislocations, and grain boundaries far removed from the ideal lattice topologies. This property renders our method safe against overzealous denoising, thus retaining meaningful disordered features in input structures.

{\color{blue}
\begin{algorithm}
\caption{Denoising process} \label{alg:denoise}
\begin{algorithmic}
    \Require{Perturbed structure $\mathbf{x}$ with atomic coordinates $\mathbf{r}$ and auxiliary information $\mathbf{z}$, and a pre-defined number of iterations $k$
    }
    \Repeat
        \State $\mathbf{r} \leftarrow \mathcal{D}(\mathbf{x}) := \mathbf{r} - \epsilon_{\theta}(\mathbf{r}, \mathbf{z})$
    \Until{convergence or $k$ is reached}
\end{algorithmic}
\end{algorithm}
}

Trained with purely synthetic data, our denoiser is applied to a wide variety of MD-perturbed systems: 
(1) BCC, FCC, and HCP Cu simulated above the melting point $T_\mathrm{m}$, as well as liquid/melt Cu perturbed around $T_\mathrm{m}$; 
\replaced{(2)}{(6)} the recently published ``DC3'' benchmark dataset for crystal structure identification \cite{Chung2022PRM} containing BCC, FCC, HCP, cubic diamond, simple cubic, and hexagonal diamond structures perturbed over a wide temperature range from cryogenic to above melting;
\replaced{(3)}{(2)} hard-to-detect transient crystal phases momentarily forming during solidification of Cu from melt; 
\replaced{(4)}{(3)} FCC, HCP, and BCC Cu containing point defects; 
\replaced{(5)}{(4)} BCC Ta containing complex dislocation networks and point defect clusters; 
\replaced{(6)}{(5)} BCC Ta containing grain boundaries; 
and finally (7) \ch{SiO2} polymorphs $\beta$-quartz, $\alpha$-cristobalite, and $\beta$-cristobalite. 
In (1) \replaced{and (3)--(6)}{--(5)}, the denoiser is shown to reduce or eliminate thermal noise, making it trivial to identify the underlying crystal structures while not destroying meaningful disordered features such as point defects, dislocations, and grain boundaries. In \replaced{(2)}{(6)}, followed by \replaced{classifiers such as a-CNA and PTM}{a-CNA classification}, the denoiser achieves perfect classification accuracies in all systems \added{at $T_\mathrm{m}$}. In (7), the generalizability of the denoiser to multi-element complex materials is validated. The results for each case study are detailed below.

\subsection{Denoising FCC, HCP, BCC, and liquid/melt Cu}
\begin{figure*}[ht]
    \centering
    \includegraphics[width=0.95\textwidth]{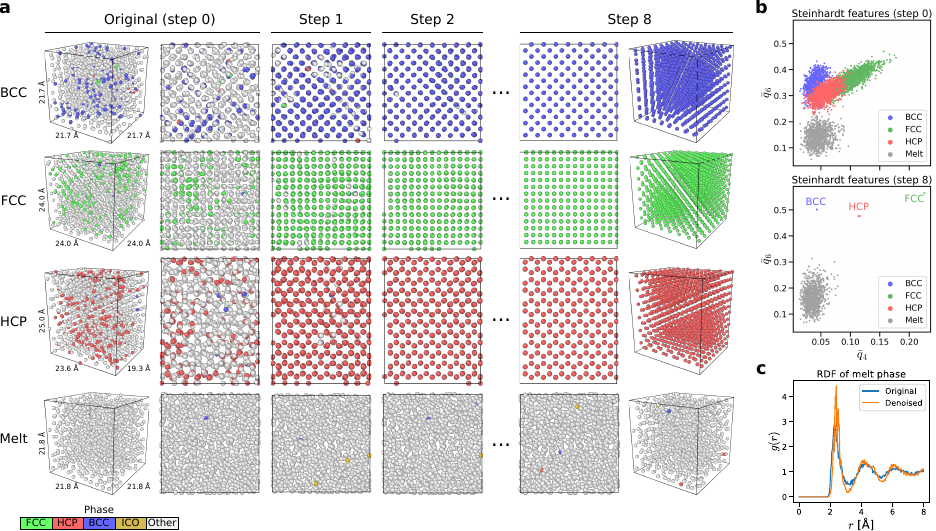}
    \caption{Iterative application of the denoiser to small cells of BCC, FCC, HCP, and liquid/melt Cu.
    \textbf{(a)} Visualization of the structures along the denoising iterations.
    \textbf{(b)} Steinhardt features $\bar{q}_4$ and $\bar{q}_6$ \cite{Lechner2008JCP} before and after the denoising.
    \textbf{(c)} Radial distribution function of the melt phase before and after the denoising.
    In \textbf{(a)}, the structures are shown in ortho (orthogonal) views along densely packed crystallographic directions, with additional perspective 3D views for steps 0 and 8. The atoms are colored according to a-CNA prediction implemented in OVITO \cite{stukowski2009visualization}. The solid phases and the melt have been annealed at 3400 K and 3000 K, respectively. ICO stands for icosahedral coordination.}
    \label{fig:denoise-bulk}
\end{figure*}

The first demonstration focuses on denoising solid BCC, FCC, and HCP crystals perturbed by thermal vibrations at above the melting point (3400 K), and liquid/melt Cu at around the melting point (3000 K), as shown in Fig.~\ref{fig:denoise-bulk}a. Before denoising, the popular adaptive CNA (a-CNA) algorithm \cite{Stukowski2012MSMSE-review} classifies most of the solid atoms (82, 77, and 77\% in BCC, FCC, and HCP, respectively) as disordered, i.e. not belonging to any of the three crystal lattice types. After just one iteration of denoising, the number of misclassified atoms is significantly reduced (14, 5, and 3\% for BCC, FCC, and HCP, respectively). The following iterations, typically within 5--8 steps, remove the remaining minor perturbations. The denoised solids resemble perfect FCC, HCP, and BCC lattices, thus trivializing subsequent phase classification. The Steinhardt order parameters $\bar{q}_4$ and $\bar{q}_6$ \cite{Lechner2008JCP} computed before and after denoising confirm that virtually all thermal perturbations imparted on the solids are removed (Fig.~\ref{fig:denoise-bulk}b). Note that after denoising, the Steinhardt quantities for BCC/FCC/HCP appear to be a single point, but are in fact about 1,000 points (each point corresponds to an atom) overlapped together.

Interestingly, denoising the Cu melt phase leaves nearly all atoms to remain disordered, as indicated by a-CNA labeling all such atoms as \textit{other} or unknown (Fig.~\ref{fig:denoise-bulk}a),  even though the atomic displacements over the denoising iterations are roughly the same as that for the solid phases (Supplementary Fig.~2). Additionally, the very first peak of the radial pair distribution function (RDF) becomes sharper and splits after denoising, and the peaks at the medium-range distances also become slightly sharper (Fig.~\ref{fig:denoise-bulk}c). Further analyses (Supplementary Fig.~3) leave precise effect of the denoiser on melt structure uncertain and perhaps deserving further scrutiny in future work. The fact that the Cu melt phase remains disordered even after denoising can be a useful property of our model not explicitly learned from its training data, which consists of only perfect and randomly distorted but otherwise ordered crystal lattices. In simulations involving solid-liquid coexistence, an example of which will be shown \replaced{later (Section~\ref{sec:cu-precip})}{next}, we certainly wish our model to denoise only the thermally distorted crystal lattices while leaving truly disordered phases disordered.

\subsection{Classification accuracy on the DC3 benchmark dataset}
\added{
Recently Chung et al. \cite{Chung2022PRM} proposed a state-of-the-art machine-learned crystal classification method, the data-centric crystal classifier (DC3). The authors provided a benchmark dataset that spans a wide variety of chemical composition, crystal structure types, and temperatures ranging from cryogenic to melting. Compared to other existing methods \cite{Honeycutt1987JPC-CNA, larsen2016robust, larsen2020revisiting, ackland2006applications, lazar2015topological, nguyen2015identification}, DC3 was shown to achieve the best accuracy on most systems in the dataset. Since the dataset includes three more crystal types, cubic diamond (\textit{cd}), hexagonal diamond (\textit{hd}), and simple cubic (\textit{sc}) in addition to FCC, BCC, and HCP, we trained a separate denoiser on all six reference structures available in the DC3 dataset. This denoiser accepts systems of at most 2 elements to accomodate the NaCl binary system (hydrogens are absent in the \ch{H2O} structures in the benchmark). We summarize in Table \ref{tab:benchmark} the improved classification accuracy at the melting $T_\mathrm{m}$ from classifying denoised structures with a-CNA and PTM, as well as the performance of DC3 \cite{Chung2022PRM} and other existing methods \cite{Honeycutt1987JPC-CNA, larsen2016robust, larsen2020revisiting, ackland2006applications, lazar2015topological, nguyen2015identification} tested in the DC3 paper. The most dramatic improvement can be seen in a-CNA, rising from near-bottom 15--57\% accuracy to on par with the state-of-the-art DC3 after just one single denoising step. With eight denoising iterations, the accuracy of a-CNA and PTM converges to the perfect classification score of 100\% on all but Fe snapshots (Supplementary Fig.~6). Confirmed by the DC3 authors, some of the Fe snapshots contain Frenkel pairs that formed spontaneously from intense thermal fluctuations at $T_\mathrm{m}$. Therefore our denoising approach indeed reaches perfect classification accuracy for BCC iron by revealing the unsuspected point defects. The difference between the performance from the a-CNA backend classifier and that from PTM on BCC Fe, namely 99.7\% and 99.9\%, can be attributed to the different ways the classifiers define atoms as defective or unknown.
}

\begin{table*}
    \centering
    \caption{
    Accuracy comparison between our classification approach (in bold font), the DC3 classifier \cite{Chung2022PRM}, and the methods tested in the DC3 work on the benchmark dataset therein. The accuracy value, computed as the fraction of correctly labeled atoms, is shown in percentage. In our denoising approach, the RMSD (root-mean-square-deviation) cutoff for PTM is 0.1. Missing entries imply non-applicable structure types for the chosen classifier.
    }
    \footnotesize
    \begin{tabular}{l r r r r r r r r r r}
    \toprule
        & Al (fcc) & Fe (bcc) & Ti (hcp) & Si (cd) & \ch{H2O} (hd) & NaCl (sc) & Ar (fcc) & Li (bcc) & Mg (hcp) & Ge (cd) \\
    \midrule
    a-CNA                        &  50.3 &  39.9 &  15.7 &       &       &       &  57.1 &  34.4 &  47.3 &       \\
    PTM$^a$                      &  95.9 &  84.3 &  82.8 &  99.9 & 100.0 &  94.6 &  96.9 &  83.1 &  95.7 &  99.2 \\
    \textbf{Denoiser (1 step) + a-CNA}  &  98.5 &  96.9     &  81.7 &       &       &       &  98.6 &  96.9 &  97.1 &       \\
    \textbf{Denoiser (1 step) + PTM}    & 100.0 &  99.5     &  98.7 & 100.0 & 100.0 &  99.3 & 100.0 &  99.6 &  99.9 & 100.0 \\
    \textbf{Denoiser (8 steps) + a-CNA} & 100.0 &  99.7$^b$ & 100.0 &       &       &       & 100.0 & 100.0 & 100.0 &       \\
    \textbf{Denoiser (8 steps) + PTM}   & 100.0 &  99.9$^b$ & 100.0 & 100.0 & 100.0 & 100.0 & 100.0 & 100.0 & 100.0 & 100.0 \\
    DC3$^a$                      &  96.9 &  86.8 &  89.4 &  99.0 &  99.2 &  95.6 &  97.5 &  85.8 &  97.4 & 100.0 \\
    i-CNA$^a$                    &  68.5 &  56.6 &  27.9 &       &       &       &  75.7 &  51.6 &  64.9 &       \\
    AJA$^a$                      &  66.9 &  35.6 &  42.4 &       &       &       &  74.0 &  34.1 &  67.0 &       \\
    VoroTop$^a$                  &  24.0 &  61.2 &  57.5 &       &       &       &  23.1 &  57.2 &  57.4 &       \\
    Chill+$^a$                   &       &       &       &  99.8 &  98.8 &       &       &       &       & 100.0 \\
    \bottomrule
    \end{tabular}
    $^a$Taken from the DC3 paper \cite{Chung2022PRM}. \\
    $^b$There is some trace amount of point defects in the BCC Fe snapshots at $T = T_m$ (Supplementary Fig.~6b).
    \label{tab:benchmark}
\end{table*}

\begin{figure}
    \centering
    \includegraphics[width=0.45\textwidth]{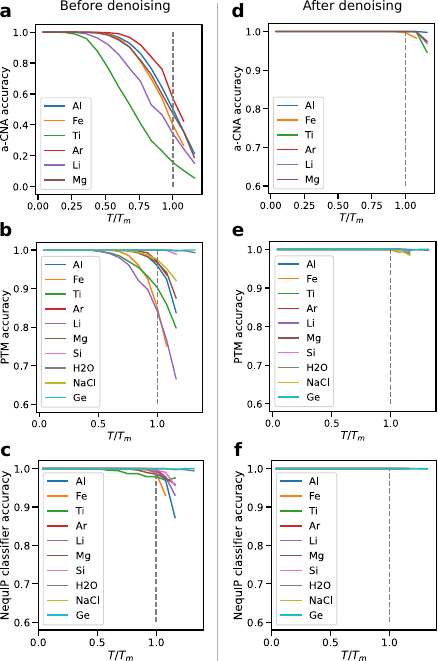}
    \caption{
    Classification performances of a-CNA, PTM, and an in-house NequIP classifier, on the DC3 benchmark dataset over a wide temperature range, are significantly improved after denoising.
    \textbf{(a, b, c)} a-CNA, PTM, and NequIP classification accuracies from nearly 0K to above melting.
    \textbf{(d, e, f)} Classification accuracies of the same models over the same temperature range, after denoising is applied.
    The melting point ($T = T_m$) is indicated by the dashed line.
    }
    \label{fig:denoise-dc3}
\end{figure}

\added{
While Table~\ref{tab:benchmark} lists the performances of various methods only at the melting point, Fig.~\ref{fig:denoise-dc3}a,b,d,e shows, over a wide temperature range, significant improvement in classification accuracy of a-CNA and PTM before and after denoising. Importantly, after denoising, both a-CNA, a common, ubiquitous classifier that is noticeably less performant, and the more updated and sophisticated PTM have essentially the same ideal performance up to the melting point, and close to perfect performance even at temperatures far above $T_\mathrm{m}$. A non-exhaustive inspection of structures above $T_\mathrm{m}$ reveals increasing defects, which explain the ostensible dip in accuracy beyond melting. Overall, the results shown so far demonstrate the utility of the denoiser for further improving existing phase classifiers and minimizing a wide range of thermal perturbations.
}

\added{
We performed further ablation tests on our approach to assess how much of the improvement could be attributed to the classifier method or to the score-based denoising idea. Using the same NequIP architecture, we trained a separate classifier based on the DC3 dataset (training details in the Methods section). What differentiates the Nequip-based denoiser and classifier is their training labels. The former was trained to predict the applied displacement vectors, which are continuous, unlimited, and information-rich targets. In comparison, the classifier had to match the structure class labels, which are discrete, far less informative, and limited to the training set. As shown in Fig.~\ref{fig:denoise-dc3}c, the classifier alone is reasonably accurate and retains more than 90\% accuracy even above melting, which is very similar to the performance of the state-of-the-art DC3 method \cite{Chung2022PRM}. Fig.~\ref{fig:denoise-dc3}f shows that even this high accuracy can be further improved to a ``perfect'' 100\% well above melting by denoising. We conclude from Fig.~\ref{fig:denoise-dc3} that score-based denoising does play a decisive role in  achieving a very high level of accuracy for classifiers of various sophistication: the classification task was made almost trivial by denoising. Here we make a cautionary note about data-driven ML classifiers trained with labels. The NequIP classifier is obviously overconfident by predicting perfectly ordered solids well above melting in Fig.~\ref{fig:denoise-dc3}f, as point defects increasingly emerge at high temperatures as discussed previously (and shown in Supplementary Fig.~6). Such defects represent rare but incorrect labels that can mislead classifiers to ignore the defects during training. Mislabeling is a major headache for supervised learning methods and hard to systematically detect and correct in large training sets. In contrast, our self-supervised denoiser does not rely on class labels, and was able to reveal, combined with non-data-driven a-CNA or PTM, increasing existence of defects, as seen in Fig.~\ref{fig:denoise-dc3}d,e.
}

\subsection{Denoising Cu solidification trajectory} \label{sec:cu-precip}
\begin{figure*}
    \centering
    \includegraphics[width=0.95\textwidth]{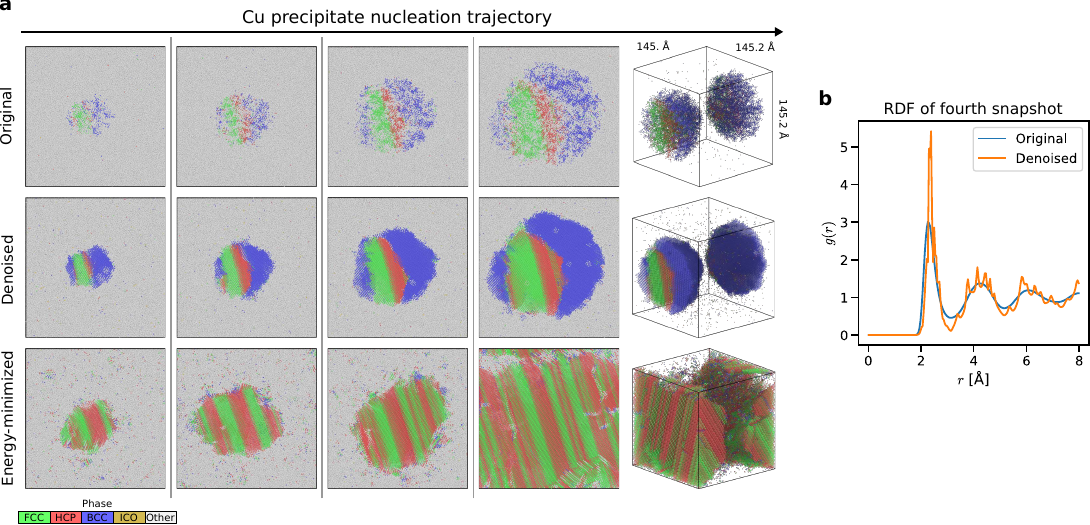}
    \caption{Denoising a dynamic trajectory of Cu solidification (314,926 atoms) significantly improves (a-CNA) phase classification without introducing unreasonably short interatomic distances.
    \textbf{(a)} Four consecutive snapshots along the trajectory, in original, denoised, and energy minmized states, are shown in ortho views with an additional perspective view for the fourth snapshot.
    \textbf{(b)} RDF of the fourth snapshot before and after denoising.
    In the perspective view, the atoms classified as \textit{other} by a-CNA are rendered transparent.
    }
    \label{fig:denoise-cu-precip}
\end{figure*}

Our denoiser is further tested here on an MD trajectory of Cu solidification from melt. Previously studied by Sadigh, et al. \cite{Sadigh2021PNAS}, solid nuclei appearing in the initial transient stages of Cu solidification are polymorphic, containing BCC, FCC, HCP, and disordered melt phase simultaneously. Characterization of such a complex transient behavior is challenging and presents a useful test case for our method. As shown in Fig.~\ref{fig:denoise-cu-precip}a, denoising four transient configurations of the trajectory results in drastic improvement in subsequent phase classification by the a-CNA algorithm. This improvement is manifested in much denser, correctly classified labels (FCC, HCP, and BCC) on the atoms within the ordered solid nucleus, as well as in considerably sharper boundaries between the phases. Here again the atoms in the disordered melt remain disordered as their a-CNA labels remain largely unchanged. Notably, a few atoms labeled as \textit{other} (i.e., unknown) are observed within the sharply defined crystal phases even after denoising. These ``unknown'' atomic motifs are likely point defects, e.g. vacancies and interstitials, as will be discussed in greater detail in the next section. Also, they appear mostly in the BCC phase likely due to the metastability of the BCC phase in Cu entropically stabilized under the high pressure (70 GPa) of the simulation. Otherwise the same BCC phase of Cu is only marginally metastable as manifested in the appearance of soft modes in its phonon spectrum. At high temperatures close to melting, such soft modes may well result in some of the atoms within the BCC phase to significantly deviate from their ideal lattice positions, resulting in formation of point defects.

It can be informative to compare denoising to steepest descent energy minimization (EM), another common method for filtering out thermal vibrations. In stark contrast to denoising, EM not only reduces the thermal perturbation but also greatly changes the nucleus structure beyond recognition (third row of Fig.~\ref{fig:denoise-cu-precip}a). Namely, under EM the solid nucleus grows considerably larger, and the transient BCC phase disappears in favor of the more stable FCC and HCP phases. The striking difference between EM and denoising can be attributed to their contrasting assumptions. The EM mapping clearly favors solid phases of lowest ground state energy and may overzealously nudge atoms towards such phases, exactly as it happens in the considered example. The denoising process, from a complementary and purely geometric perspective, relies on \replaced{less biased}{an unbiased}, equal prior probabilities of the reference phases included in training.

To further investigate whether denoising introduces unwanted or unphysical artifacts, in Fig.~\ref{fig:denoise-cu-precip}b we plot the RDF of the last snapshot from Fig.~\ref{fig:denoise-cu-precip}a. The RDF of the denoised structure generally matches that of the original. Having been trained to reduce thermal perturbations, the denoiser likely has also learned not to bring atoms to excessively short distances of each other (that would be unphysical). The sharp peaks of the denoised RDF are attributed to the three ordered crystal phases, each of which contributes its own discrete set of sharply defined interatomic distances. 

Note that the denoiser does in fact ``denoise'' some atoms within the melt phase into local environments regarded as crystalline (BCC/FCC/HCP) by the a-CNA classification, as evidenced by the increased number of solid labels in the melt region after denoising (Fig.~\ref{fig:denoise-cu-precip}a). The appearance of such ``crystalline'' atoms reflects that even in fully disordered liquid, statistically a small fraction of the atomic motifs may momentarily resemble a crystal phase. Our denoiser then acts locally and further enhances such resemblance, thus making a-CNA (a strictly local classifier itself) to recognize such atoms as crystalline. Observing that these misclassified atoms are few and far isolated, we accept this minor artifact as a worthwhile trade-off. A more elaborate classifier, perhaps accounting for local environments beyond just the nearest neighbors, should be able to cleanly separate the ordered and the disordered melt phases in such instances.

\subsection{Denoising FCC, HCP, and BCC Cu containing point defects}
\begin{figure}
    \centering
    \includegraphics[width=0.45\textwidth]{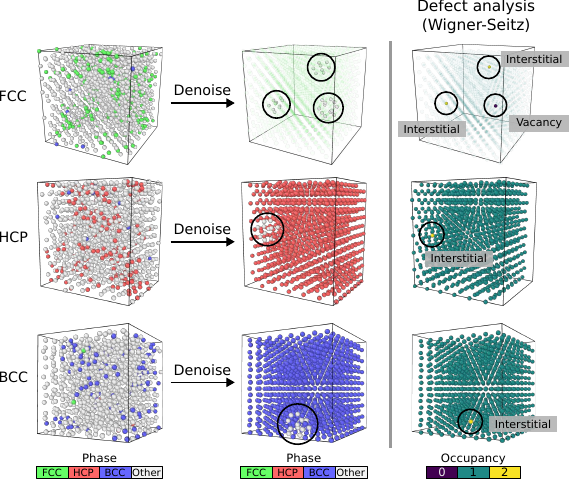}
    \caption{Denoising facilitates point defect identification in thermally perturbed FCC, HCP, and BCC Cu crystals. Point defects completely obscured by thermal vibrations are cleanly revealed after denoising (circled in black). When applied to the denoised structures, the Wigner-Seitz defect analysis (OVITO) correctly assigns mass content to each crystal defect: site occupancies 0, 1, and 2 correspond to a vacancy, a regular atom, and an interstitial, respectively. After denoising, the regular atoms in the FCC crystal (top row) are rendered semi-transparent to more clearly reveal the point defects.}
    \label{fig:denoise-point-defects}
\end{figure}

As an example of point defect characterization, the denoiser is applied to FCC, HCP, and BCC Cu each containing an intentionally inserted extra atom followed by annealing in MD at 3400 K. These structures were denoised into ideal lattices with local regions of disorder unknown to a-CNA (Fig.~\ref{fig:denoise-point-defects}). The Wigner-Seitz defect analysis (OVITO) confirms that these regions of disorder indeed correspond to point defects. Notably, for the FCC crystal, the thermal vibrations were sufficiently intense to spontaneously generate two more point defects, a Frenkel pair of one vacancy and one interstitial.

The example in Fig.~\ref{fig:denoise-point-defects} demonstrates desirable outcomes of denoising crystal structures containing point defects. Although the denoiser aims to modify local atomic motifs towards ideal topology, it cannot do so on regions of point defects simply due to extra or missing atoms. In such a case, the model appears to not significantly alter the local topologies around the defects while denoising the rest of the bulk into an ideal lattice.

\subsection{Denoising BCC Ta containing dislocations} \label{sec:Ta_disloc}
\begin{figure*}
    \centering
    \includegraphics[width=0.95\textwidth]{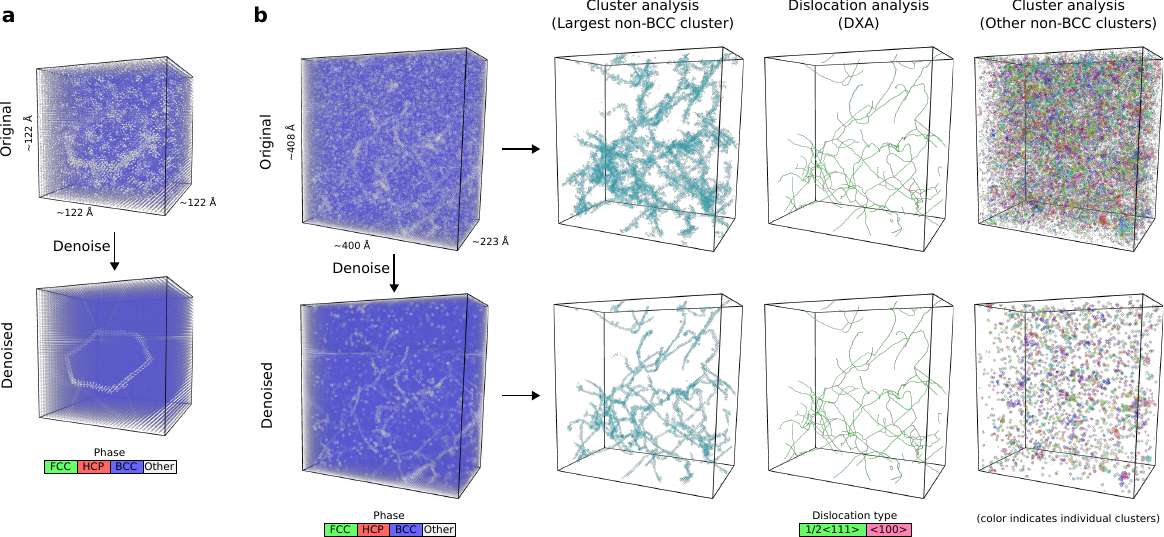}
    \caption{Denoising BCC single crystal Ta containing dislocations (and point defects).
    \textbf{(a)} A single dislocation loop annealed at 2500 K.
    \textbf{(b)} A snapshot of a relatively large MD simulation (1.97 million atoms) of single crystal Ta undergoing deformation at 2000 K, which results in a network of entangled dislocation lines and a large number of point defect clusters. The DXA method for dislocation analysis (OVITO) confirms the presence of dislocations and delineates the dislocation network topology. The cluster analysis (OVITO) aims to separate the point defect clusters from the dislocations. The atoms are normally colored by a-CNA prediction except for the cluster/dislocation analysis, in which the BCC atoms are rendered transparent.
    }
    \label{fig:denoise-dislocations}
\end{figure*}

Our model is similarly effective for denoising \added{structures containing} lattice dislocations. For a toy example, a hexagon-shaped dislocation loop inserted into BCC Ta was annealed at 2500 K ($0.8 T_m$) and subsequently denoised (Fig.~\ref{fig:denoise-dislocations}a). Again, satisfyingly, the denoiser does not significantly alter local atomic configurations near the dislocation loop while cleanly denoising the surrounding crystal bulk.

Note that although the denoiser was trained on ideal and noised Cu lattices, it is applicable to Ta or any other elemental crystal of FCC, BCC or HCP lattice structure. This transferability is achieved by simply scaling the input (noised) structure to match the interatomic distance of the corresponding Cu phase. The output (denoised) structure would then be re-scaled back to its original dimensions. 

As a more realistic and difficult test, the denoiser was applied to help reveal a complex dislocation network in a BCC Ta crystal subjected to plastic deformation at 2000 K (Fig.~\ref{fig:denoise-dislocations}b). Similar to the case of the single dislocation loop, the dislocation network (as colored by a-CNA) is more sharply defined after denoising. Subsequent application of the DXA algorithm \cite{Stukowski2012MSMSE-DXA} to the original and the denoised configurations results in nearly identical dislocation networks, which testifies to the exceptional robustness of DXA against thermal perturbation, and confirms that the dislocations are better captured by a-CNA after denoising.


Despite DXA's already high performance in dislocation characterization, the denoiser still benefits or complements DXA by facilitating the characterization of the point defect clusters that were either left as debris in the wake of dislocation motion or produced by dragging jogs formed at dislocation intersections \cite{Stimac2022}. Focusing on non-BCC atoms clustered within a cutoff distance of 3.2 Å, a large cluster (corresponding to the dislocation network) and a high concentration of small clusters are observed. Without denoising, the small clusters may simply be manifestation of noise based on visual interpretation. However, after denoising, the small clusters resemble and likely capture the point defect debris. This is unlikely a case of the denoiser failing to denoise non-dislocation regions into perfect lattice for two reasons: (1) the presence of the point defect debris is known \textit{a priori}, and (2) the denoiser clearly denoises the non-dislocation region in the toy case of the single dislocation loop, with virtually no point defects left (Fig.~\ref{fig:denoise-dislocations}a).

\subsection{Denoising BCC Ta containing grain boundaries} \label{sec:Ta_GB}
\begin{figure}
    \centering
    \includegraphics[width=0.45\textwidth]{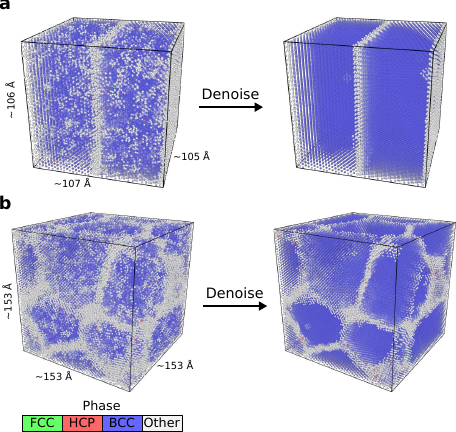}
    \caption{The denoising process helps reveal grain boundaries obscured by thermal noise.
    \textbf{(a)} A bi-crystal of BCC Ta (64,000 atoms) containing two planar grain boundaries.
    \textbf{(b)} A polycrystal of BCC Ta (187,921 atoms) containing a network of grain boundaries. Both crystals were annealed in MD simulations at 2500 K which caused minor coarsening in the polycrystal. After denoising, trace amount of point defects becomes visible in both examples. The BCC atoms are rendered slightly semi-transparent.}
    \label{fig:denoise-grains}
\end{figure}

To test how our method performs on crystals containing grain boundaries, the denoiser was applied to a Ta bi-crystal containing two tilt boundaries. Prior to denoising, the bi-crystal was annealed at a high temperature of 2500 K. As shown in Fig.~\ref{fig:denoise-grains}a, denoising does not alter the topology of the defects, and results in two near-perfect BCC crystals separated by two perfectly planar grain boundaries, with trace amount of point defects likely emitted from the boundaries into the bi-crystal interior. A more complex test case is shown in Fig.~\ref{fig:denoise-grains}b, where a polycrystal consisted of 12 grains of BCC Ta had been similarly annealed at 2500 K. Here again, denoising removes thermal noise while still revealing a few point defects in the grain interiors.

\subsection{Denoising \ch{SiO2} polymorphs}
\begin{figure*}
    \centering
    \includegraphics[width=0.9\textwidth]{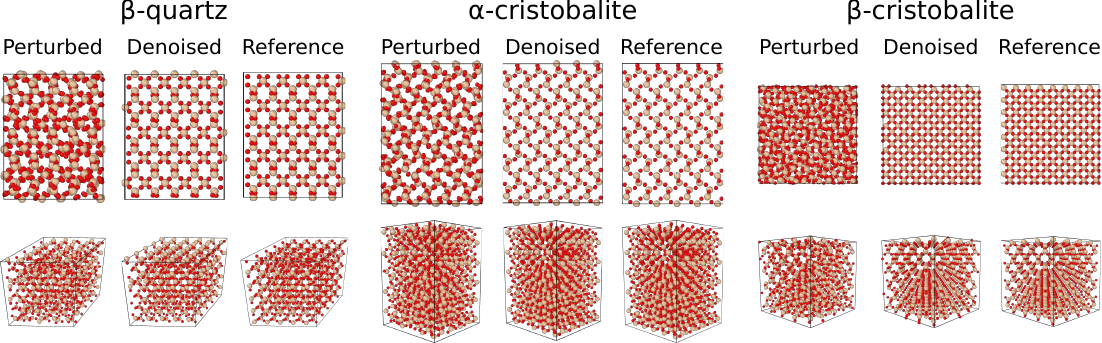}
    \caption{A separately trained denoiser effectively removes thermal perturbations (simulated at 1000 K) in \ch{SiO2} $\beta$-quartz, $\alpha$-cristobalite, and $\beta$-cristobalite polymorphs. The denoised structures match the ideal references used during training. 16 iterations were used for denoising. The silicon atoms are colored in beige, and the oxygens in red.}
    \label{fig:denoise-sio2}
\end{figure*}

Finally, to validate the generalizability of our approach, we trained a separate denoiser for minimizing thermal perturbation in \ch{SiO2} systems. Similar to the demonstration shown in Fig.~\ref{fig:denoise-bulk}, the denoiser was applied to MD-perturbed high-temperature silica polymorphs $\beta$-quartz, $\alpha$-cristobalite, and $\beta$-cristobalite (Fig.~\ref{fig:denoise-sio2}). Although we also trained the denoiser with reference $\alpha$-quartz topology, the inherent similarity between $\alpha$- and $\beta$-quartz complicates the task of generating stable and distinctive MD-perturbed snapshots of $\alpha$-quartz compared to $\beta$-quartz. Therefore $\alpha$-quartz is not included in this preliminary result. Regardless, the denoiser removes virtually all thermal perturbation in the silica polymorphs. Future work may include extending to larger multi-element systems in the presence of disordered features.

\section{Discussion}
In developing our new method we have taken to a geometric and statistical perspective on delineating order-disorder features in MD simulations of solids, a problem particularly difficult at elevated temperatures approaching. We regard two seemingly distinct tasks of classifying ordered phases and locating disordered defects as fundamentally the same problem, which is addressed by denoising. Based on a statistical score function, the denoiser presented and tested in this work effectively reduces and minimizes thermal noise in ordered solids without impacting isolated disordered defects and the liquid/melt phase. \added{Combined with various classifiers, it was able to achieve perfect classification accuracy in a public benchmark dataset over existing methods. Ablation tests against no or single-step denoising as well as supervised classifier based on the same network architecture provided more evidence for the effectiveness of our approach.} To support our conclusions, in the preceding sections we applied denoising to reveal the underlying ``anomaly'' structures across the entire spectrum of crystal disorder, namely (0D) point defects, (1D) dislocation lines, (2D) grain boundary, and (3D) liquid phases, all distinct from the ordered phases that the denoiser was trained with. \added{For future work, more comprehensive and quantitative effort is necessary to validate the merit of the denoiser for aiding defect characterization.} For denoising BCC/FCC/HCP, our model was trained only once and only on the three ordered structures but then shown to successfully but not excessively denoise the much more complex structures used for testing. We relate this useful ability of not overly denoising defective atoms to the inductive bias of our graph network model with its limited number of message-passing steps. Trained entirely on synthetic data, our model is not derived from any deep physical insights about topology and geometry of the reference structures and, as such, does not require careful data curation. Thus, its extension to other ordered structures should be straightforward. 

Our method holds unique advantages (and disadvantages) over two other methods widely used for reducing thermal perturbation: energy minimization (EM) and vibration-averaging (VA). With the right parameters and done over a small number of iterations, EM can lead to similar results as that of the denoiser. However, over many iterations, EM may grow or shrink certain phases due to the intrinsic bias associated with the use of interatomic potential. An example of EM being overzealous and distorting configurations beyond recognition with respect to the original state is in Section~\ref{sec:cu-precip}. The purely geometric denoiser, on the other hand, does not require a known or developed interatomic potential. Further, the denoiser is unbiased in the sense that the reference lattices used in training have equal prior probability. Although the denoising graph model is more expensive than VA (which entails simple averaging operations), VA can potentially smear out atomic motion by averaging over a time interval. Such a distortion can be fairly significant since, to average out thermal vibrations, VA requires time averaging intervals of hundreds or even thousands of time steps. Our denoising method, on the other hand, treats every time snapshot separately and does not coarse-grain over time. As is often the case, no one method used for denoising is singularly superior to all other existing methods. We hope that our approach finds its own place among existing and emerging methods for structure and defect classification, and can serve as an accurate and efficient pre-processing filter to facilitate application of more computationally demanding methods of structural analysis such as DXA. 

Based on an equivariant graph network model architecture, our denosing model is readily extendable to more complex reference structures and materials by incorporating additional information such as atom types into the graph embedding. In addition to extending the method to chemically complex systems, our ongoing and future efforts may also focus on its computational efficiency and scalability. Finally, beyond atomic structures, the ability of our model to achieve state-of-the-art classification accuracy through iterative denoising score functions suggests the idea may be useful in other disciplines for enhanced accuracy and/or robustness against fluctuations.

\section{Methods}

\subsection{Theoretical justification of the denoising model} \label{sec:theory}
Our approach to denoising builds on the theory of statistical learning of score functions \cite{Vincent2011scorematching} that establishes equivalence between denoising and score matching. Consider a probability distribution function $q(\mathbf{x})$ that exists in principle but is analytically intractable due to the high dimensionality of the data space $\mathbf{x} \in \mathbb{R}^d$ ($d \gg 1$). Focusing on approximating the gradient of the log-probability density $\nabla_{\mathbf{x}} \log q(\mathbf{x})$---also known as the score function \cite{Hyvarinen2005}---rather than $q(\mathbf{x})$ itself circumvents the often intractable problem of finding the normalization constant for $q(\mathbf{x})$. Score matching then amounts to finding an approximating model $s_{\theta}(\mathbf{x})$ with parameters $\theta$ to match the score function, with the \textit{score matching } loss \cite{Hyvarinen2005}
\begin{equation}
    L_{\mathrm{SM}} = 
    \frac{1}{2} \mathbb{E}_{q(\mathbf{x})} 
    \left[\left\Vert
        s_{\theta}(\mathbf{x}) - \nabla_{\mathbf{x}} \log q(\mathbf{x})
    \right\Vert^2 \right].
\end{equation}
Nevertheless, the term $\nabla_{\mathbf{x}} \log q(\mathbf{x})$ is still unknown. To address this, consider approximating $q(\mathbf{x})$ by adding isotropic Gaussian noises of variance $\sigma^2$ to the (clean) data samples $\mathbf{x}$, resulting in noised samples $\mathbf{x}' =\mathbf{x} +\sigma \epsilon$, where $\epsilon \sim \mathcal{N}(0, \mathbf{I})$, and the approximating distribution
\begin{equation}
q_{\sigma}(\mathbf{x} '|\mathbf{x}) = Z e^{- \Vert \mathbf{x} '-\mathbf{x}\Vert ^{2}/2\sigma ^{2}},
\end{equation}
where $Z$ is a normalization constant. This way, instead of the original loss, we minimize the \textit{denoising score matching} loss based on the key insight from Ref.~\cite{Vincent2011scorematching} to train with pairs of clean and corrupted data points:
\begin{equation}
    L_{\mathrm{DSM}} = 
    \frac{1}{2} \mathbb{E}_{q_{\sigma}(\mathbf{x}', \mathbf{x})} 
    \left[\left\Vert
        s_{\theta}(\mathbf{x}') - \nabla_{\mathbf{x}'} \log q_{\sigma}(\mathbf{x}' | \mathbf{x})
    \right\Vert^2 \right],
\end{equation}
where the new score function $\nabla_{\mathbf{x}'} \log q_{\sigma}(\mathbf{x}' | \mathbf{x})$ can be computed via
\begin{align}
\log q_{\sigma}(\mathbf{x} '|\mathbf{x}) & =\log( Z) -\frac{1}{2\sigma ^{2}}\Vert \mathbf{x} '-\mathbf{x}\Vert ^{2}, \nonumber \\
\nabla _{\mathbf{x}'} \log q_{\sigma}(\mathbf{x} '|\mathbf{x}) & =-\frac{1}{\sigma ^{2}}(\mathbf{x} '-\mathbf{x})  = \frac{1}{\sigma ^{2}}(\mathbf{x} -\mathbf{x}') = -\frac{\epsilon }{\sigma },
\end{align}
revealing that the score function points from noisy samples $\mathbf{x}'$ to clean ones $\mathbf{x}$. This observation also implies that learning the score function is equivalent to training a denoising model. To better see this, note that the denoising score matching loss can now be simplified into
\begin{equation} \label{eq:dsm_loss_simple}
    L_{\mathrm{DSM}} = 
    \frac{1}{2} \mathbb{E}_{q_{\sigma}(\mathbf{x}', \mathbf{x})} 
    \left[\left\Vert
        s_{\theta}(\mathbf{x}') + \frac{\epsilon}{\sigma}
    \right\Vert^2 \right].
\end{equation}
After scaling equation~(\ref{eq:dsm_loss_simple}) by a factor of $\sigma$ and incorporating a noise prediction model $\epsilon_{\theta}(\textbf{x}')$ that aims to predict the applied noise, then the loss function can be written as
\begin{equation} \label{eq:dsm_loss_final}
    L_{\mathrm{DSM}} = 
    \frac{1}{2} \mathbb{E}_{q_{\sigma}(\mathbf{x}', \mathbf{x})} 
    \left[\left\Vert
        \epsilon - \epsilon_{\theta}(\mathbf{x}')
    \right\Vert^2 \right],
\end{equation}
finally establishing the connection between the score function model $s_{\theta}(\textbf{x}')$ and the denoising model $\epsilon_{\theta}(\textbf{x}')$ by $\epsilon_{\theta}(\textbf{x}') = - \sigma s_{\theta}(\textbf{x}')$. This clarifies the meaning of $\epsilon_\theta$ in \replaced{Algorithm~\ref{alg:denoise}}{Equation~1}: it is the scaled score function defining the noise added to clean data. While the noise amplitude $\sigma$ is a hyper-parameter, it can be estimated by a fitted/trained denoising model itself from a noisy input. Hereafter, we refer to $\epsilon_\theta$ as a score function for brevity. Along the above steps for connecting between score matching and denoising, we have omitted certain details for brevity. For a rigorous formulation, see \cite{Vincent2011scorematching}.

The ideal score function is a gradient field in the data space that converges to clean data points used to train the denoiser. For better intuition, a toy score function is visualized in Supplementary Fig.~1, which illustrates that following the score function is the same as denoising. In our context where ideal FCC, HCP and BCC lattices were used for training, a perturbed (noised) input configuration may be denoised into one of the three ideal (clean) structures that it resembles the most. At the same time, a highly perturbed input configuration bearing no resemblance to any of the three ideal reference configurations is unlikely to be meaningfully denoised resulting in unknown or divergent values of predicted noise. The case studies considered in this work all suggest that our denoising model does not significantly alter such disordered structures, including melt, point defects, dislocations, and grain boundaries. This property is instrumental in allowing the denoising model to reveal underlying crystalline order without impacting meaningful disordered features in thermally perturbed configurations.

The score function plays a central role in modern likelihood-based generative models such as the denoising diffusion probabilistic model (DDPM) \cite{Sohl-Dickstein2015-DPM, Ho2020-DDPM} and score-based generative model \cite{Song2020-unified}, which can be unified under the same framework \cite{Song2020-unified}. Among its numerous recent achievements \cite{Yang2022-dpm-review}, DDPM has been applied to crystal and molecular structure generations \cite{Xie2022-dpm-crystal, Xu2022-dpm-molecule}. In this work, however, we apply the score matching method for denoising rather than generative applications, and focus on a limited number of reference crystal structures instead of many (thousands or millions) training images/structures.

\subsection{Model training}
Our (clean) data samples are reference crystal structures of interest represented by the atomic coordinates $\mathbf{r}$ and the auxiliary information $\mathbf{z}$: $\mathbf{x}^{(i)} \rightarrow (\mathbf{r}^{(i)},\mathbf{z}^{(i)})$. The noise prediction model $\epsilon_{\theta}$ was trained with entirely synthetic data \added{(Algorithm~\ref{alg:train})}, which is generated by adding Gaussian noises to the atomic coordinates $\mathbf{r}' = \mathbf{r} + \sigma \epsilon$, with $\sigma \sim \mathcal{U}(0, \sigma_{\max})$ drawn uniformly up to $\sigma_{\max} \approx 13\%$ of the shortest interatomic distance, adhering to Lindemann's law on mean-squared thermal displacement of solids before melting \cite{Lindemann1910}. Since thermal displacements in classical MD are sums of phonon modes that follow Boltzmann distributions rather than i.i.d. Gaussians, our working hypothesis is that the correlated thermal fluctuations have fewer degrees of freedom than i.i.d. Gaussians, and a model trained with the latter can adequately handle the former. Our implemented loss function, slightly adjusted from equation~(\ref{eq:dsm_loss_final}), is
\begin{equation} \label{eq:loss}
    L = \mathbb{E}_{\mathbf{r},\mathbf{z}, \sigma, \epsilon} \left[\Vert \sigma \epsilon -\epsilon _{\theta }(\mathbf{r} + \sigma \epsilon,\mathbf{z})\Vert ^{2}\right].
\end{equation}

\begin{algorithm}
\caption{Denoiser training} \label{alg:train}
\begin{algorithmic}    
    \Require{Ideal reference dataset $\mathbf{D}$, initial model parameters $\theta$, maximum noise amplitude $\sigma_{\max}$, gradient descent optimizer Optim, and learning rate $\eta$
    }
    \Repeat
       \State Sample $\mathbf{x} \sim \mathbf{D}$
       \State Sample $\sigma \sim \mathcal{U}(0, \sigma_{\max})$
       \State Sample $\epsilon \sim \mathcal{N}(0, \mathbf{I})$
       \State $(\mathbf{r}, \mathbf{z}) \leftarrow \mathbf{x}$
       \State $L \leftarrow \mathbb{E}_{\mathbf{r},\mathbf{z}, \sigma, \epsilon} \left[\Vert \sigma \epsilon -\epsilon _{\theta }(\mathbf{r} + \sigma \epsilon,\mathbf{z})\Vert ^{2}\right]$
       \State $\theta \leftarrow \mathrm{Optim}(L, \theta, \eta)$
    \Until{convergence}
\end{algorithmic}
\end{algorithm}

The idea of mixing training data with random noises is not new in either general-purpose or scientific machine learning. Adding a small amount of noise to the training data, known sometimes as the ``noise trick'', is a well-established data augmentation or regularization technique in general-purpose machine learning to reduce overfitting and increase model robustness \cite{JMLR:v11:vincent10a}. For example, Zhou et al. \cite{Zhou2014PRL, Zhou2019PRB} adopted a hybrid training data pipeline of MD trajectories and Gaussian noise displacements to fit the potential energy surface of crystalline solids. A similar method was used by Chung et al.\ \cite{Chung2022PRM} to identify ordered solid phases. The noise trick was also adopted to train GNN surrogate models for physical simulations \cite{Pfaff2020, Sanchez-Gonzalez2020}. This work, however, makes denoising noise-corrupted inputs the centerpiece rather than merely a regularization technique.


The model was trained with randomly drawn $\mathbf{x} \in \{$FCC, BCC, HCP$\}$, using the AdamW optimizer \cite{loshchilov2017decoupled} and a learning rate of $2\times 10^{-4}$, over 20,000 weight updates in minibatches of 32 samples. Each FCC/HCP/BCC cell consists of roughly 1000 atoms. The training was carried out using PyTorch \cite{paszke2019pytorch} and PyTorch-Geometric \cite{fey2019fast}. All other training parameters, if unspecified in this work, default to values per PyTorch 1.11.0 and PyTorch Geometric 2.0.4.

\subsection{Equivariant graph network model (NequIP)}
The denoising model output is a vector (on each atom) that should be equivariant under translation, rotation and mirror operations---the same requirements for force fields or interatomic potentials. We adopted a customized version of the E(3)-equivariant NequIP model \cite{e3nn, Batzner2021NC-nequip}, which guarantees such equivariance. NequIP is primarily built upon the idea of equivariant tensor product between two inputs of irreducible representations, or irreps, resulting in another irreps as the output. Unlike regular tensor products, the tensor products in NequIP are parametrized by learnable weights and are therefore termed \textit{WeightedTP} in this work. Since the exact mathematical details of the equivariant tensor product can be dense and complex, we refer to the original work for their precise description \cite{Batzner2021NC-nequip}. We chose to directly predict vector outputs (noising displacements) rather than a scalar output.

\begin{figure}
    \centering
    \includegraphics[width=0.45\textwidth]{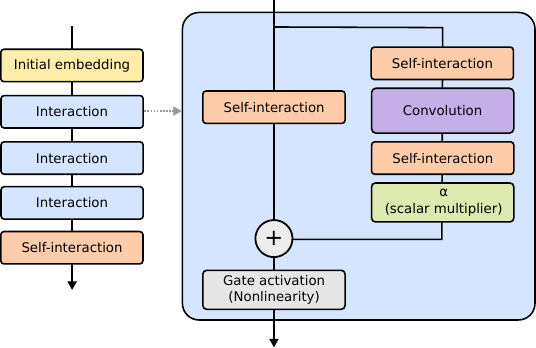}
    \caption{The NequIP model architecture.}
    \label{fig:nequip}
\end{figure}

The main components of our NequIP variant consist of the initial embedding, the interaction layers, and the final self-interaction layer to produce the noise output (Fig.~\ref{fig:nequip}).

In the initial embedding, the structure input is converted to an atomic graph, with $\mathbf{h}_i$ as some attributes for node or atom $i$; $\tilde{\mathbf{h}}_{i}$ as another set of attributes for the same node $i$; and $\mathbf{e}_{ij}$ as the vector for the \emph{directed} edge from node $i$ to node $j$. Transformed from atom type information by a trainable embedding matrix, $\mathbf{h}_i$ initially only holds scalar information ($l = 0$, where $l$ is the tensor rank or the degree of representation) but is typically expanded to hold information of higher tensor ranks ($l = 1, 2, ...$) in subsequent layers. $\tilde{\mathbf{h}}_{i}$, also transformed from atom type information by a trainable embedding matrix, holds only scalar information and does not change throughout the model layers. The embedding matrix is not needed for the single-element systems but necessary for the \ch{SiO2} polymorphs. We used Atomic Simulation Environment \cite{larsen2017atomic} and PyTorch-Geometric \cite{fey2019fast} for the conversion to graphs.

Each interaction layer consists of several sub-operations: the self-interaction, the convolution, SkipInit \cite{de2020batch}, and the gate activation. The self-interaction updates the attributes of each node $\mathbf{h}_i$ via the WeightedTP operation with $\tilde{\mathbf{h}}_{i}$ and does not aggregate information from neighbor nodes:
\begin{equation}
    \mathbf{h}'_{i} =\mathrm{WeightedTP}\left(\mathbf{h}_{i} ,\tilde{\mathbf{h}}_{i}\right)
\end{equation}
The convolution updates the attributes of each node $\mathbf{h}_i$ as the sum of the WeightedTP operations between the neighboring nodes $\mathbf{h}_{j}$ and the spherical harmonics of neighboring edges $Y(\hat{\mathbf{e}}_{ij})$, with the weights learned from the edge distances $\Vert \mathbf{e}_{ij}\Vert$ via a multilayer perceptron (MLP):
\begin{equation}
    \mathbf{h}'_{i} =\frac{1}{Z} \sum_{j \in N(i)}\mathrm{WeightedTP}_{\Vert \mathbf{e}_{ij}\Vert }(\mathbf{h}_{j}, Y(\hat{\mathbf{e}}_{ij}))
\end{equation}
where $\hat{\mathbf{e}}_{ij}$ is the normalized version of $\mathbf{e}_{ij}$ and is therefore a unit vector pointing from node $i$ to node $j$, $N(i)$ denotes the neighbor nodes of node $i$, and $Z$ is a normalization constant. The MLP contains one hidden layer. The initial layer of the MLP is the basis function values expanded from the edge distance. For the SkipInit mechanism \cite{de2020batch}, the scalar multipliers $\alpha$ are learned from yet another WeightedTP operation between $\mathbf{h}_i$ and $\tilde{\mathbf{h}}_{i}$ (similar to the self-interaction operation). The gate activation applies equivariant nonlinearities \cite{weiler20183d} to the node attributes.

In the end, the final self-interaction layer serves to transform the node attributes $\mathbf{h}_i^{(L-1)}$ from the second to last layer $L-1$, which may hold scalar, vectorial, and tensorial features at node $i$, into a single vector describing the noise output:
\begin{equation}
    \mathbf{h}^{(L)}_i
    = (\epsilon_{\theta})_i
    = \mathrm{WeightedTP}\left(\mathbf{h}^{(L-1)}_{i} ,\tilde{\mathbf{h}}^{(L-1)}_{i}\right),
\end{equation}

The complexity of the model is largely determined by the specified irreps format for the node and edge attributes. For example, an array of 4 scalars and 8 vectors can be written as 4x0e + 8x1o, with the numbers 4 and 8 describing the multiplicities, the numbers 0 and 1 describing the tensor rank, and the letters e (even) and o (odd) describing the parity. Higher multiplicities and tensor ranks can often result in better performance but also larger memory and computational requirements. We intentionally kept the model complexity small in favor of scalability to structures of millions of atoms. The model settings are listed in \replaced{Supplementary Table~1}{Table~2}.

\subsection{NequIP classifier}
\added{
The NequIP-based classifier was trained on 90\% of the Al, Fe, Ti, Si, \ch{H2O}, and NaCl snapshots from the DC3 benchmark dataset, with the rest being the validation dataset. Specifically, for each of the Al, Fe, Ti, Si, \ch{H2O}, and NaCl systems, there are ten snapshot files per temperature increment, and nine were taken for training, with the remaining one for validation. In other words, this classifier was not trained on any of the Ar, Li, Mg, and Ge structure data from the DC3 dataset. Applying the classifier to Ar, Li, Mg, and Ge amounts to scaling the unit cell dimensions to match that of the corresponding training data with the same structure type. For example, Ge is slighter larger than Si, and therefore is scaled down (roughly 90\%) before feeding to the classifier. The model parameters of the NequIP classifier are listed in Supplementary Table 1.
}

\subsection{Molecular dynamics simulations}
This section describes the MD simulations used to demonstrate the capabilities of our denoising method. These are simulations of (1) BCC, FCC and HCP Cu structures, both defect-free and with point defects; (2) a solid crystal nucleus growing inside melted Cu; (3) crystal plasticity in single crystal of Ta in the BCC phase; (4) Ta grain boundaries; and (5) \ch{SiO2} in $\beta$-quartz and cristobalite polymorphs. All simulations were performed with periodic boundary conditions using  LAMMPS \cite{LAMMPS}. 

The MD simulations for Cu using the embedded-atom method (EAM) potential by Mishin {\it et al.} \cite{MishinPRB2001} were performed in BCC, FCC and HCP cells containing 1024, 1372, and 1152 atoms respectively. Since the bulk BCC phase is dynamically unstable in Cu with imaginary phonon frequencies at ambient conditions, the calculations were performed at a pressure of 60 GPa, where the BCC phase becomes metastable.

Melting points of the three phases at 60 GPa have been calculated to be 3030~K for BCC, 3066~K for HCP, and 3073~K for FCC \cite{Sadigh2021PNAS}. Although FCC remains the thermodynamically stable and thus preferred phase below 3073~K, free energies of three phases are very close under these pressure and temperature conditions. At slightly higher pressures (71.6 GPa and 85 GPa), the phase diagram of the model of Cu contains triple points where two of the three solid phases and the liquid phase coexist \cite{Sadigh2021PNAS}. We have taken advantage of these thermodynamic proximity of three crystal phases and the melt to set up an MD simulation of a polymorphic critical solid nucleus simultaneously containing all three solid phases surrounded by melt. The simulation contained 314,928 Cu atoms and was initiated in an isobaric-isoenthalpic (NPH) ensemble at 70 GPa from a small near-equilibrium nucleus with coexisting FCC and HCP ordered regions containing about 200 FCC atoms and 300 HCP atoms, respectively. Upon switching to an isobaric-isothermal (\replaced{NPT}{NVT}) ensemble at the same pressure and temperature 2800~K, the solid nucleus grows and partially transforms to the BCC phase. 

Interatomic interactions in tantalum were modeled using a well-known EAM potential developed by Li et al. \cite{Li03}. For simulations of crystal plasticity in Ta described in Section \ref{sec:Ta_disloc}, the crystals were created by arranging atoms in a BCC lattice within a cubic or an orthorhombic periodic supercell with repeat vectors aligned along the cube axes of the BCC lattice. Dislocations were seeded into the crystals in the form of one or several hexagon-shaped prismatic loops of the vacancy type, following the procedure introduced in \cite{Zepeda17}. For the configuration in Fig.~\ref{fig:denoise-dislocations}a, a single dislocation loop was introduced at the center of a cube-shaped simulation box made of 101,853 atoms and annealed at temperature 2000 K. The complex network of dislocations shown In Fig.~\ref{fig:denoise-dislocations}b was generated by initially introducing 12 randomly positioned dislocation loops into a $\sim 2$ million atoms box, annealing the model at 2500 K and zero pressure and then subjecting the crystal to uniaxial compression along the [001] crystallographic axis at a ``true'' strain rate of $2 \times 10^8$/s for 2 ns while maintaining pressure near zero in an NPH ensemble.

For Ta grain boundaries in Section \ref{sec:Ta_GB}, the periodic bi-crystal containing two $\Sigma 5 (310)$ symmetric tilt grain boundaries was created by joining two crystal blocks of different lattice orientations obtained by rotating two half-crystals in the opposite directions along the common $\langle 100 \rangle$ tilt axis. The Ta polycrystal was assembled using atomsk \cite{hirel2015atomsk} from 12 randomly seeded grains. Both the bi-crystal and the polycrystal were annealed at 2500 K and zero pressure.

For silica, the Tersoff potential developed by Munetoh et al. \cite{munetoh2007interatomic} was adopted. Simulations were performed in the NVT ensemble with unit cell parameters taken from the Encyclopedia of Crystallographic Prototypes \cite{mehl2017aflow, hicks2019aflow, hicks2021aflow}, and with temperature ramping from 600 K to 2000 K in steps of 200 K every 10 ps. 5x5x5, 5x5x5, and 4x4x4 supercells were used for the $\beta$-quartz (space group 181), $\alpha$-cristobalite (92), and $\beta$-cristobalite (227), respectively.
\section*{Acknowledgements}
TH and FZ acknowledge support by the Critical Materials Institute, an Energy Innovation Hub funded by the U.S.\ Department of Energy, Office of Energy Efficiency and Renewable Energy, and Advanced Materials and Manufacturing Technologies Office. BS, CP, NB and VB are partially supported by the Laboratory Directed Research and Development (LDRD) program (22-ERD-016) at Lawrence Livermore National Laboratory. This work was performed under the auspices of the US Department of Energy by Lawrence Livermore National Laboratory under contract No. DE-AC52-07NA27344. JC was partially supported by the Department of Mechanical Engineering’s startup grant at Boston University. We thank Dr.\ R.\ Freitas for providing the DC3 benchmark dataset and Dr.\ A.\ Stukowski for useful discussions. Computing support for this work came from the LLNL Institutional Computing Grand Challenge program.

\section*{Author Contributions}
TH implemented the denoising graph network model and analyzed the denoised MD configurations. 
BS, NB, and CP generated the MD data to be denoised. FZ supervised the research with inputs from all authors.

\section*{Data Availability}
\replaced{
All data required to reproduce this work, namely the molecular dynamics snapshots (ideal reference, thermally perturbed, and denoised) can be requested from Tim Hsu and Fei Zhou. Some small, simple structure data for demonstration purposes are included in the code repository of this work (see Code Availability). The DC3 dataset used for benchmarking can be found at \url{https://github.com/freitas-rodrigo/DC3}.
}
{All data required to reproduce this work can be requested from the corresponding author(s).}

\section*{Code Availability}
The source code, the trained denoiser \added{and classifier} models, and a \deleted{simple} demo \replaced{for denoising simple structures}{for this work} are available at \url{http://www.github.com/llnl/graphite}.

\section*{Competing interests}
The authors state that there is no conflict of interest.

\printbibliography
\onecolumn
\newpage
\setcounter{figure}{0}
\renewcommand{\figurename}{Supplementary Figure}
\renewcommand{\theHfigure}{Supplement.\thefigure}

\setcounter{table}{0}
\renewcommand{\tablename}{Supplementary Table}
\renewcommand{\theHtable}{Supplement.\thetable}

\section*{Supplementary Information}

\subsection*{Visualization of toy score function} \label{si-sec:toy-score-fn}
\begin{figure}[h]
    \centering
    \includegraphics[width=0.8\textwidth]{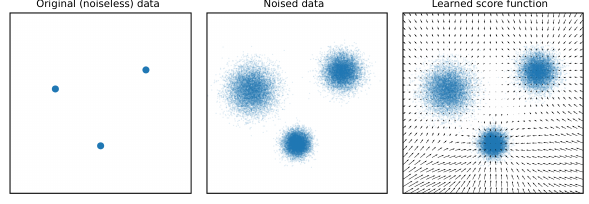}
    \caption{Visualization of a toy score function. Given a noisy dataset based on some clean data samples, the score function for the noisy data is a gradient field (indicated by the arrows) that converges to the original noiseless data points. The score function shown here is not exact, but was learned/estimated with a simple multilayer perceptron using PyTorch.}
    \label{si-fig:toy-score-fn}
\end{figure}

Here we briefly provide intuition and visualization for a score function based on a toy data example. Suppose the data space is only two-dimensional. Let there be three (noiseless) data points noised with Gaussians of different standard deviations. Then the score function for the noised data is a gradient field in the 2D space that converges on the three original data points (Supplementary Fig.~\ref{si-fig:toy-score-fn}). Therefore, the act of denoising is equivalent to iteratively following the gradient field that eventually points to the clean data, which in this work is analogous to the ideal reference atomic motifs such as BCC, FCC, and HCP.

\newpage
\subsection*{The effect of denoising on melt/liquid Cu} \label{si-sec:denoise-liq}
Since the denoising model was trained to denoise perturbed solid structures (namely, BCC, FCC, and HCP), how the model behaves with disordered melt/liquid structure input is unknown. Here we investigate the effect of the denoising process on the melt/liquid Cu phase.

\begin{figure}
    \centering
    \includegraphics[width=0.4\textwidth]{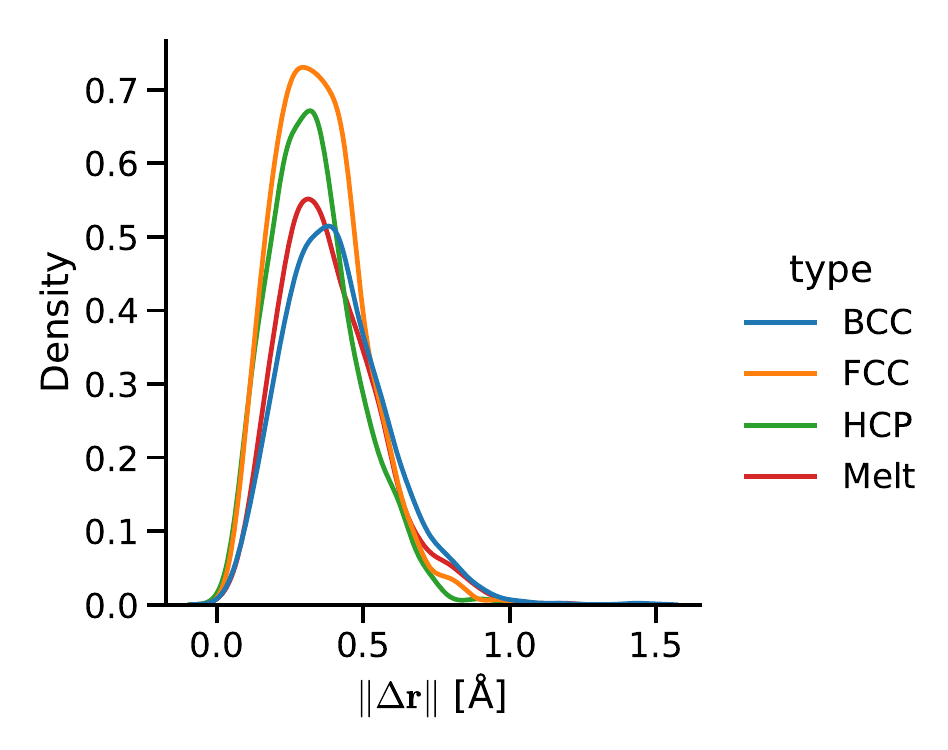}
    \caption{Histogram of atomic displacement magnitudes from denoising BCC, FCC, HCP, and melt Cu. This displacement data is based on the snapshots from Fig.~2, specifically the difference in atomic coordinates between the original state (step 0) and the denoised state (step 8).}
    \label{si-fig:liq-disp}
\end{figure}

From denoising BCC, FCC, HCP, and melt/liquid Cu shown in Fig.~2, the atomic displacements of the melt phase is similar to those of the solids, indicating that all atoms move over roughly the same distance range regardless of the phase (Supplementary Fig.~\ref{si-fig:liq-disp}). This is an interesting observation, as the melt phase remains largely disordered after denoising according to the a-CNA algorithm and the measured RDF, even after the same level of displacement as that of the solid phases. However, the melt Cu structure in Fig.~2 is relatively small and may not provide enough statistics for a sufficiently smooth RDF. We therefore applied denoising to a larger cell of melt Cu phase (256,000 atoms) and measured the RDF, as shown in Supplementary Fig.~\ref{si-fig:denoise-liq-rdf}. Similar to the result shown in Fig.~2c, the general RDF pattern after denoising is similar to that of the original, with peak splitting in the short range and more pronounced ordering in the medium range. We initially hypothesized that the short range peaks would match those of ideal BCC, FCC, and HCP structures. However, from Supplementary Fig.~\ref{si-fig:denoise-liq-rdf}c, it is unclear whether these short range peaks closely match those of the solid phases, with some peaks that seemingly overlap with those of the solid phases and some peaks that clearly do not.

\begin{figure}
    \centering
    \includegraphics[width=0.5\textwidth]{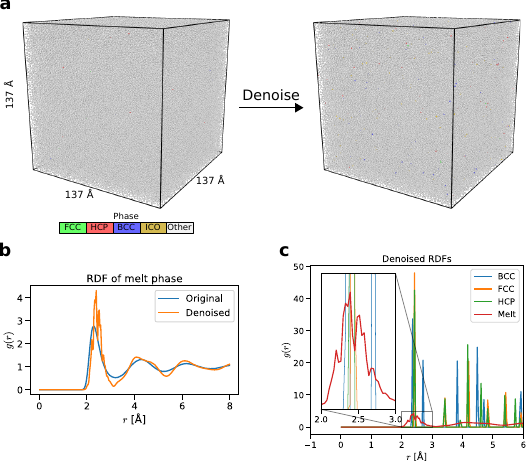}
    \caption{Comparison of the RDF of a denoised melt Cu configuration with that of the original structure and those of BCC, FCC, and HCP Cu. \textbf{(a)} Visualization of the original and denoised configurations where the atoms are colored by the a-CNA prediction. \textbf{(b)} Comparison between the original and denoised configurations. \textbf{(c)} Comparison between the denoised RDF of the melt phase and those of the solid phases (BCC, FCC, and HCP).}
    \label{si-fig:denoise-liq-rdf}
\end{figure}

Therefore, how the denoising model topologically impacts the disordered liquid/melt phase is not yet fully understood. We posit that the emergence of the short range peaks and the medium range ordering are attributed to the atoms being moved into equidistant positions from the nearest neighbors as a result of the model learning to form ideal atomic motifs. In other words, the model may be aiming to denoise liquid structures into ideal lattices, but only to a very limited extent. It is also possible that the melt phase, though highly disordered, is associated with some inherent structure or order \cite{ganesh2006signature}, which in turn influences the behavior of the denoising model. In any case, a more thorough investigation would be necessary to uncover the precise effect of denoising on melt/liquid structures. However, such an inquiry is outside of the scope of this work and is left as another dedicated future effort related to the inherent order and symmetry in liquid.

\newpage
\subsection*{Denoised trajectory of BCC Ta plastic deformation}
\begin{figure}
    \centering
    \includegraphics[width=0.9\textwidth]{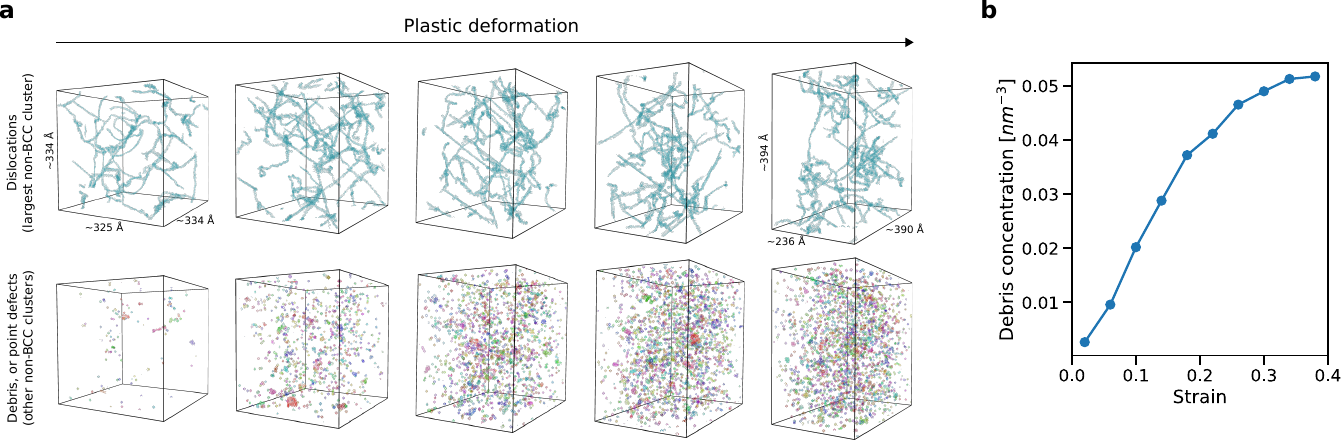}
    \caption{Denoising helps reveal that debris (point defects) concentration increases during plastic deformation as a result of moving dislocations. \textbf{(a)} The dislocation and the debris are elucidated via the same analysis applied to Fig.~6b in the main text, namely, with denoising followed by cluster analysis of non-BCC atoms. \textbf{(b)} Quantitative measurement of the debris concentration as a function of strain. The debris concentration is measured as the number of non-BCC clusters (excluding the dislocations) divided by the unit cell volume. Whether an atom belongs to BCC or other phases is determined by a-CNA prediction.}
    \label{si-fig:plastic-deformation}
\end{figure}

The plastic deformation trajectory prior to the snapshot shown in Fig.~6b is also denoised and analyzed here. Using the same analysis for Fig.~6b, the dislocations and the debris (point defects) are elucidated. This analysis confirms the movements of the dislocations, as well as the corresponding increase in the debris concentration (Supplementary Fig.~\ref{si-fig:plastic-deformation}), which is known to increase under plastic deformation. Therefore, our denoising model is shown to help capture complex networks of dislocations and point defects (or debris) in perturbed and deformed structures that are otherwise difficult to analyze. Indeed, analysis of point defects evolution in original (noisy) structures generally cannot yield meaningful results as a large number of atoms are misclassified as defective only due to thermal displacements.

\newpage
\subsection*{Denoising vs. vibration-averaging}
\begin{figure}
    \centering
    \includegraphics[width=0.7\textwidth]{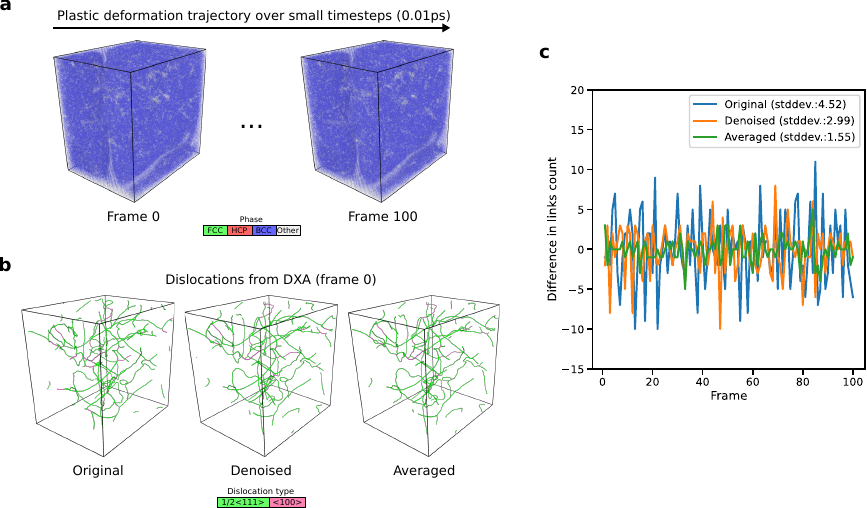}
    \caption{Comparison between the effects of denoising and vibration-averaging on the frame-by-frame dislocation network topological changes based on a short segment of the plastic deformation trajectory studied in this work.
    \textbf{(a)} The trajectory made of 100 frames (with 0.01 ps timesteps) with the atoms colored by a-CNA prediction.
    \textbf{(b)} The dislocation networks at the initial frame, extracted for the original, denoised, and averaged trajectories. 
    \textbf{(c)} The frame-by-frame difference in the number of network links along the (original, denoised, and averaged) trajectory. Since the topological changes are hard to visually discern, the difference in the number of network links is measured and plotted.}
    \label{si-fig:denoise-vs-avg}
\end{figure}

Besides energy minimization, vibration-averaging is another common technique for reducing thermal perturbations. Here we briefly provide a comparison analysis between denoising and averaging for the plastic deformation trajectory mentioned in Fig.~6b, with an emphasis on how denoising/averaging impacts the frame-by-frame dislocation topological changes.
Evolution of dislocation networks in terms of local topological events can provide key insights into fundamental dislocation mechanisms at the atomic scale, but their detailed analysis is made extremely difficult and generally muddled by incessant, spurious flickers caused by thermal vibrations.
Focusing on the subtle changes in the dislocation network topology, 100 frames of the plastic deformation trajectory were simulated over small time steps of 0.01 ps (Supplementary Fig.~\ref{si-fig:denoise-vs-avg}). For a given frame, the vibration-averaging was done by averaging the atomic coordinates from the past 100 frames (including the current frame).

Visually, dislocation configurations extracted with DXA from the trajectories display numerous visible but subtle changes in their topology in going from one frame to the next. To quantify these changes, we count the frame-by-frame difference in the number of dislocation links (dislocation lines connecting two network nodes, i.e. edges of the dislocation network) along the trajectory (Supplementary Fig.~\ref{si-fig:denoise-vs-avg}c). For the original (noisy) configurations, this number has a standard deviation of 4.52. However, it is unclear which changes correspond to ``true'' topological events and which are flickers induced by thermal fluctuations. In vibration-averaged configurations, the corresponding standard deviation is greatly reduced (1.6). In contrast, the difference in the number of links varies by a standard deviation of 3.0 across for the denoised configurations. This higher value may suggest that the vibration-averaging technique is overzealous and eliminate true topological events as a consequence of slightly smearing out the dynamics. Our denoising approach is immune to such dynamic distortions. Thus, this example illustrates how denoising can serve as an accurate pre-processing filter for methods of structural analysis such as DXA.

\newpage
\subsection*{Denoising the DC3 benchmark dataset}
The denoiser was applied to the public benchmark dataset introduced by Chung et al. \cite{Chung2022PRM} for crystal structure identification. The impact of the number of denoising steps on subsequent classifier accuracy at $T = T_m$ is shown in Supplementary Fig.~\ref{si-fig:denoise-dc3}.

\begin{figure}
    \centering
    \includegraphics[width=0.8\textwidth]{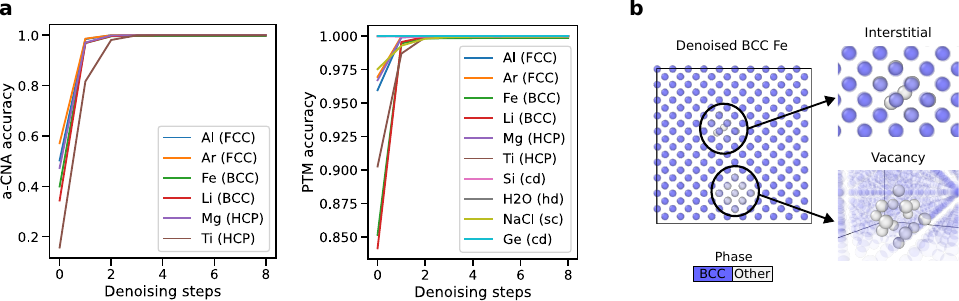}
    \caption{Findings from denoising and classifying the DC3 benchmark dataset.
    (\textbf{a}) a-CNA and PTM classification accuracies along the denoising steps.
    (\textbf{b}) Based on the denoised structures, there is some trace amount of Frenkel defect pairs in BCC Fe. One such pair (circled) is shown here.
    In (\textbf{b}), the atoms labeled as BCC (by CNA) are rendered semi-transparent.
    }
    \label{si-fig:denoise-dc3}
\end{figure}

\newpage
\subsection*{Model parameters}
The NequIP-based denoiser parameters are listed in Supplementary Table~\ref{si-tab:model-params}.

\begin{table}
    \centering
    \caption{NequIP-based denoiser model parameters. Values in parenthesis correspond to the classifier variant trained for classifying the DC3 benchmark dataset.}
    \begin{tabular}{l l l}
    \toprule
    Name & Value \\
    \midrule
    Irreps for initial node attributes $\mathbf{h}$                & 10x0e               \textit{(8x0e)} \\
    Irreps for auxiliary node attributes $\tilde{\mathbf{h}}$      & 10x0e               \textit{(8x0e)} \\
    Irreps for intermediate/hidden node attributes $\mathbf{h}$    & 10x0e + 10x1e       \textit{(8x0e + 8x4e + 8x6e)} \\
    Irreps for edge spherical harmonics $Y(\hat{\mathbf{e}}_{ij})$ & 1x0e + 1x1e + 1x2e  \textit{(1x0e + 1x4e + 1x6e)} \\
    Number of basis functions for expanding edge distance $\Vert \mathbf{e}_{ij}\Vert$ & 16 \\
    Radius cutoff for edge distance $\Vert \mathbf{e}_{ij}\Vert$ & 3.2 Å \\
    Number of hidden neurons for the MLP & 64 \\
    Normalization constant $Z$ & 12 \\
    \bottomrule
    \end{tabular}
    \label{si-tab:model-params}
\end{table}
\end{document}